\documentclass{caosp309}

\usepackage{graphicx}
\usepackage{subcaption}

\usepackage{natbib}
\articleNo{301}
\pubyear{2024}
\volume{NN}
\volnumber{NN}
\firstpage{1}
\received{MM, DD, YYYY}
\accepted{MM, DD, YYYY}

\def\BibTeX{{\rm B\kern-.05em{\sc i\kern-.025em b}\kern-.08em
             T\kern-.1667em\lower.7ex\hbox{E}\kern-.125emX}}

\begin{document}

\hauthor{S.\,Overall and J.\,Southworth}

\title{EBOP MAVEN}
\subtitle{A machine learning model for predicting eclipsing binary light curve fitting parameters}

\author{
	Stephen Overall\orcid{0009-0009-5312-6583}
	\and
	John Southworth\orcid{0000-0002-3807-3198}
}

\institute{
	Astrophysics Group, Keele University, Staffordshire, ST5 5BG, UK\\
	\email{s.p.overall@keele.ac.uk}
}

\date{November 13, 2024}
\maketitle

\begin{abstract}
Detached eclipsing binary stars (dEBs) are a key source of data on fundamental
stellar parameters. While there is a vast source of candidate systems in the
light curve databases of survey missions such as {\it Kepler} and TESS, published
catalogues of well-characterised systems fall short of reflecting this abundance.
We seek to improve the efficiency of efforts to process these data with the
development of a machine learning model to inspect dEB light curves and predict
the input parameters for subsequent formal analysis by the {\sc jktebop} code.
\keywords{binaries: eclipsing -- methods: data analysis -- methods: machine-learning}
\end{abstract}

\section{Introduction}
Detached eclipsing binary systems are a key source of stellar parameters
used in the development and testing of models of stellar evolution. By
combining photometric light curve and spectroscopic radial velocity (RV)
observations  it is possible to measure the components' physical characteristics
with great precision \citep{Andersen91aar,TorresAndersen+10aar}.

Machine learning (ML) approaches are becoming widely used in many areas of
science and technology, often as a means to gain new insights into existing
datasets or as a way of addressing the vast volumes of data produced by the
latest scientific programmes. The key feature of the machine learning approach
is that a given program learns to carry out its task through direct experience
with the data rather than through a predefined algorithm.

\section{Developing the model}
{\sc ebop maven} is a CNN (Convolutional Neural Network) model implemented in python
with the TensorFlow \citep{Abadi+2015tensorflow} and Keras \citep{Chollet+2015keras}
libraries. CNN models are a type of supervised machine learning model,
one that requires training before use, which are widely used in computer vision
scenarios. They typically consist of one or more convolutional layers which
with training learn the convolution kernels best able to extract the characteristic
features within the input data from which a subsequent neural network can make
predictions.

We have developed a regression model which views the 1-D time-series data of
a phase-folded dEB light curve to predict the value of six parameters used
as the input for subsequent formal analysis with the {\sc jktebop} code
\citep{Southworth08mnras}. The predicted parameters are the sum ($r_{\rm A}+r_{\rm B}$)
and ratio ($k\equiv r_{\rm B}/r_{\rm A}$) of the stars' fractional radii, the central
surface brightness ratio ($J$), the orbital eccentricity through the
Poincaré elements ($e\cos{\omega}$ and $e\sin{\omega}$), and the orbital
inclination through the primary impact parameter ($b_{\rm P}$).

The model was trained on a fully synthetic dataset of 250,000 labelled
phase-folded light curves, split 80:20 between training and validation
instances. The distribution of the training data was chosen to cover the
parameter space supported by {\sc jktebop}. While training, augmentation
steps added random noise and shifts to the data. Dropout layers are included
in the model as a regularisation measure to combat overfitting.

\section{Testing and results}
The test dataset consists of synthetic instances representing a population of
physically plausible dEB systems. Stellar parameters were taken from
MIST \citep{ChoiDotter+16apj} stellar models with population demographics based
on the product of the initial mass function from \cite{Chabrier03pasp} and the
binarity function from \cite{WellsPrsa21apjs}. The treatment of apparent
brightness, noise and limb darkening was based on the TESS mission.

A second test dataset of real dEB systems was assembled based on membership
of the DEBCat \citep{Southworth15debcat} catalogue of well-characterised detached
systems and the availability of TESS timeseries photometry. The test labels were
derived from the published works giving the most recent characterisation of
each system. A processing pipeline was implemented to prepare and fit light curves 
from these systems using {\sc jktebop}, and a set of control fits were made
with input parameters based on the test labels. The final set of systems
represent a range of targets where it was possible to achieve a stable control
fit with results comparable to those published.

\begin{figure}[thp]\centering
	\subcaptionbox{The predictions made for V456 Cyg (circles), V570 Per (squares)
		and AI Phe (diamonds) plotted against the corresponding label values}
		{\includegraphics[width=11.5cm,clip=]{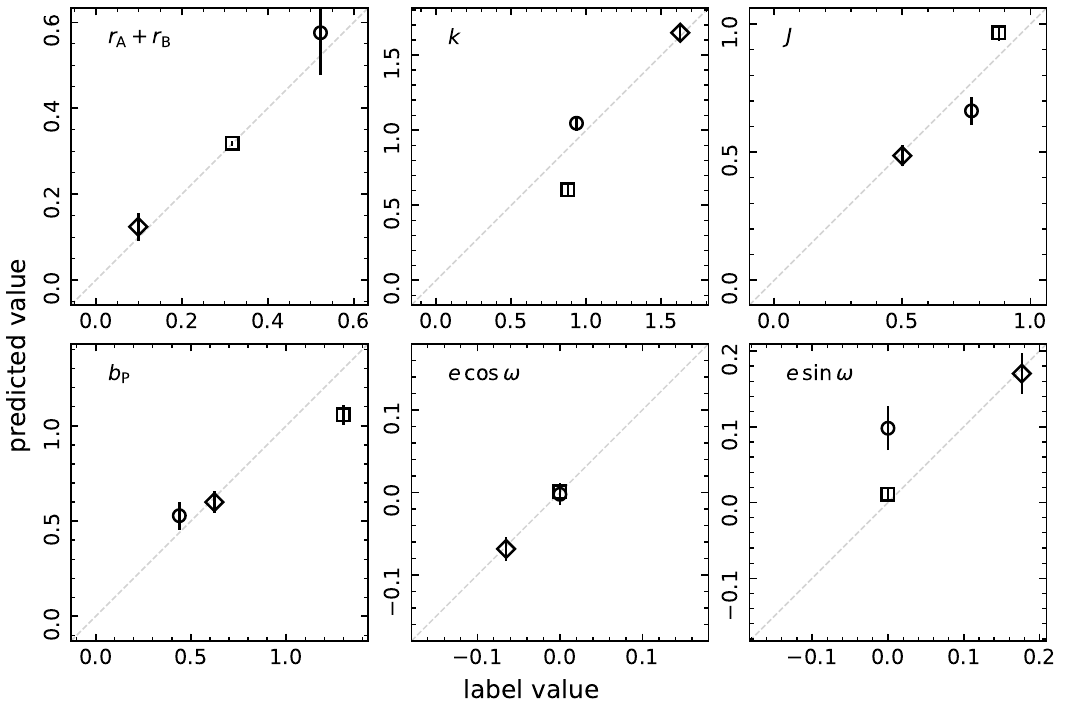}\label{fig:1a}}
		
	\vspace{0.25cm}
	
	\subcaptionbox{The phase-folded light curves of V456 Cyg, V570 Per and AI Phe
			with corresponding generated light curves based on the predicted input parameters
			(shifted by 0.3\,mag) and the fitted characterisation with {\sc jktebop}
			(shifted by 0.6\,mag)}
		{\includegraphics[width=11.5cm,clip=]{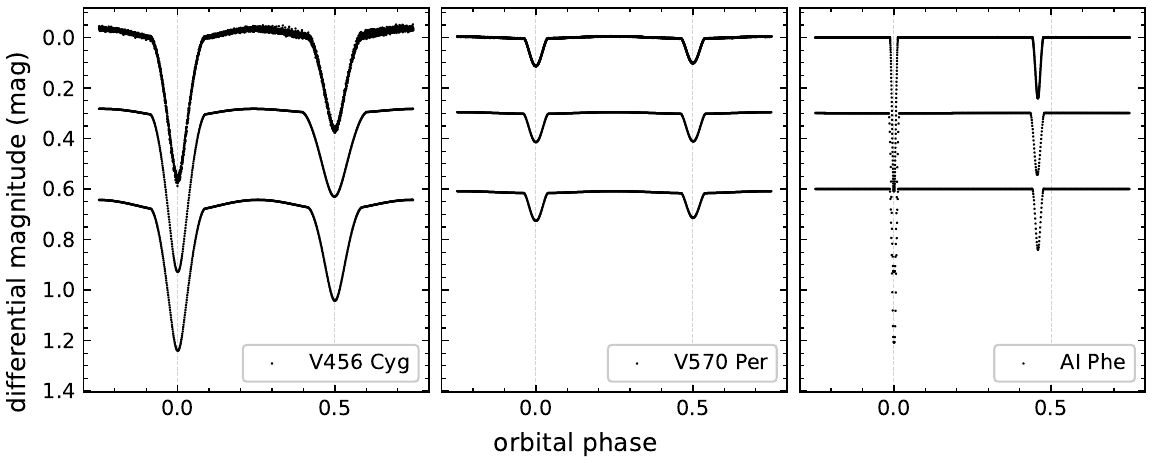}\label{fig:2a}}
			
	\caption{The {\sc ebop maven} predictions for a subset of the test dataset of
		real dEB systems below which are the corresponding phase-folded light curves
		of the input feature, and generated light curves of both the predicted and final fit}
	\label{fig:1}
\end{figure}

The {\sc ebop maven} model was used to predict the fitting parameters for the real
dEB systems. With the predictions used as the input parameters for the fitting
step of the processing pipeline, while all other settings and processes were
carried over from the control fits, 22 of the 23 systems yielded a final fitted
characterisation in agreement with the equivalent control fit. The exception was
AI Phe (see Fig.\,\ref{fig:1}), however this is an especially challenging target
where the secondary star is significantly larger than the primary ($k\simeq1.6$).

\section{Conclusion and next steps}
We have demonstrated that it is possible to train a machine learning model
to predict input parameters for dEB light curve fitting with the 
{\sc jktebop} code. When tested with TESS photometry from real systems
we achieved a good fit in 22 of 23 targets.

Development of the model is ongoing and a full article is being prepared
for publication. The code to build, train and test the version of the model
discussed here is publicly available on
GitHub\footnote{\url{https://github.com/SteveOv/ebop_maven/tree/kopal2024}}.
The {\sc ebop maven} model will be incorporated into a processing
pipeline with which a catalogue of dEB systems will be assembled.
This is intended to provide sufficient data for the selection of suitable
targets for further observation and formal characterisation.

\acknowledgements
We gratefully acknowledge the support of the Science and Technologies
Facilities Council (STFC) in the form of a PhD studentship.

\bibliography{kopal2024-fp05}
\end{document}